\documentclass[conference]{IEEEtran}
\IEEEoverridecommandlockouts
\usepackage{cite}
\usepackage{amsmath,amssymb,amsfonts}
\usepackage{algorithmic}
\usepackage{graphicx}
\usepackage{multirow}
\usepackage{float}
\usepackage{textcomp}
\usepackage{gensymb}
\usepackage{xcolor}
\def\BibTeX{{\rm B\kern-.05em{\sc i\kern-.025em b}\kern-.08em
    T\kern-.1667em\lower.7ex\hbox{E}\kern-.125emX}}
\begin{document}

\title{Breaking the Memory Wall for AI Chip 
with a New Dimension}

\author{\IEEEauthorblockN{1\textsuperscript{st} Eugene Tam}
\IEEEauthorblockA{ 
\textit{IC League Inc.}\\
Haining, China \\
eugene.tam@ieee.org}
\and
\IEEEauthorblockN{2\textsuperscript{st} Shenfei Jiang}
\IEEEauthorblockA{ 
\textit{IC League Inc.}\\
Haining, China \\
shenfei.jiang@icleague.com}
\and
\IEEEauthorblockN{3\textsuperscript{st} Paul Duan}
\IEEEauthorblockA{ 
\textit{IC League Inc.}\\
Haining, China \\
paul.duang@icleague.com}
\and
\IEEEauthorblockN{4\textsuperscript{st} Shawn Meng}
\IEEEauthorblockA{ 
\textit{IC League Inc.}\\
Haining, China \\
Shawn.meng@icleague.com}
\and
\IEEEauthorblockN{5\textsuperscript{st} Yue Pang}
\IEEEauthorblockA{ 
\textit{IC League Inc.}\\
Haining, China \\
yue.pang@icleague.com}
\and
\IEEEauthorblockN{6\textsuperscript{st} Cayden Huang}
\IEEEauthorblockA{ 
\textit{IC League Inc.}\\
Haining, China \\
cayden.huang@icleague.com}
\and
\IEEEauthorblockN{7\textsuperscript{st} Yi Han}
\IEEEauthorblockA{ 
\textit{IC League Inc.}\\
Haining, China \\
hetachy01@sina.com}
\and
\IEEEauthorblockN{8\textsuperscript{st} Jacke Xie}
\IEEEauthorblockA{ 
\textit{IC League Inc.}\\
Haining, China \\
jacke.xie@icleague.com}
\and
\IEEEauthorblockN{9\textsuperscript{st} Yuanjun Cui}
\IEEEauthorblockA{ 
\textit{IC League Inc.}\\
Haining, China \\
yuanjun.cui@icleague.com}
\and
\IEEEauthorblockN{10\textsuperscript{st} Jinsong Yu}
\IEEEauthorblockA{ 
\textit{IC League Inc.}\\
Haining, China \\
jinsong.yu@icleague.com}
\and
\IEEEauthorblockN{11\textsuperscript{st} Minggui Lu}
\IEEEauthorblockA{ 
\textit{IC League Inc.}\\
Haining, China \\
Lumingui@gmail.com}
}
\IEEEoverridecommandlockouts
\IEEEpubid{\makebox[\columnwidth]{978-1-5386-5541-2/18/\$31.00~\copyright2020 IEEE \hfill} \hspace{\columnsep}\makebox[\columnwidth]{ }}

\maketitle

\IEEEpubidadjcol

\begin{abstract}
Recent advancements in deep learning have led to the widespread adoption of artificial intelligence (AI) in applications such as computer vision and natural language processing. As neural networks become deeper and larger, AI modeling demands outstrip the capabilities of conventional chip architectures. Memory bandwidth falls behind processing power. Energy consumption comes to dominate the total cost of ownership. Currently, memory capacity is insufficient to support the most advanced NLP models. In this work, we present a 3D AI chip, called Sunrise,  with near-memory computing architecture to address these three challenges. This distributed, near-memory computing architecture allows us to tear down the performance-limiting memory wall with an abundance of data bandwidth. We achieve the same level of energy efficiency on 40nm technology as competing chips on 7nm technology. By moving to similar technologies as other AI chips, we project to achieve more than ten times the energy efficiency, seven times the performance of the current state-of-the-art chips, and twenty times of memory capacity as compared with the best chip in each benchmark.
\end{abstract}


\section{Introduction}
AI is now used widely in a diverse set of tasks, ranging from face recognition to real-time speech processing to autonomous transportation. Among these applications, deep neural networks have been responsible for the most recent AI breakthroughs, like DeepMind’s AlphaGo’s stunning defeat of the Go world champion. Google has developed a high-quality deep neural machine translation system between 17 languages \cite{b4}. Since then, natural language processing models have only grown more powerful and are now capable of programming, writing stories, and even composing poems. These powerful AI models require deeper and larger nets. In 2019, Nvidia published natural language model Megatron with 8.5 billion parameters. This year, Microsoft introduced the next generation Turing-NLG with 17 billion parameters. Recently, OpenAI released an even larger model and current state-of-the-art, GTP-3 \cite{b26}. With 174 billion parameters, GPT-3 takes 350GB memory and \$12 million to train\cite{b27}.

It is clear that newer AI models require more parameters, more memory, and larger training sets. They will only continue to grow in size and training cost. To address these future demands, we propose a new architecture for AI chip.  In this paper, we discuss two key components of this new architecture: heterogenous chip integration technology and single-form memory.

Our results demonstrate the potential of this new architecture. This chip, fabricated on a 40nm process, achieves performances comparable with current-state-of-the-art chips, which use generations more advanced fabrication processes than the 40nm. We further extrapolate the performances of our chip to current fabrication processes. Our chip is projected to be able to hold 12 billion parameters on a single chip, while current best on the market only holds 8 billion on a whole wafer.

\section{Background}
\begin{figure}[htbp]
\centerline{\includegraphics[width=\linewidth]{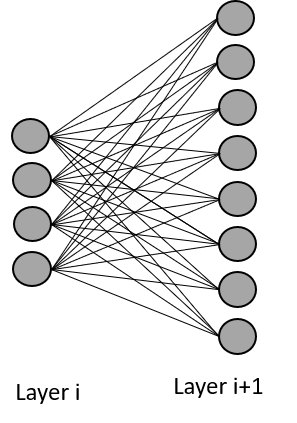}}
\caption{Fully-connected layers in a neural network.}
\label{fig:network}
\end{figure}

Neural networks consist of layers of nodes and the connections between them. Each node stores a value. During infernece, the first layer contains the input values. The values of the nodes in subsequent layers are calculated from the values in the previous layer's nodes. The last layer is the output of the neural network. Figure \ref{fig:network} shows connections between two layers.  A neural neural network may be deep (i.e. a large number of layers). It may also be wide-- each layer may have large number of nodes.
Large amounts of computing power and training data are generally regarded as the drivers of modern breakthoughs in artificial intelligence. To sustain the momentum in the field, we need to address the three challenges for AI chips: the memory wall, energy efficiency, and on-chip memory capacity.

Computational power has increased 60\% every year since the 80's, as IC fabrication technology advances with Moore’s law. Yet memory performance, namely DRAM, has increased only 7\% every year during the same time\cite{b6}\cite{b7}. The gap between processor and memory speed has been widening for more than two decades. As a result, data bandwidth cannot match the processing speed. This is the well-known “memory wall.”  The memory wall has been a challenge for CPUs for decades. As AI computations consume much more data than most other applications, the memory wall is especially relevant in AI chip design.

There are two main existing approaches to addressing the memory wall problem. One is increasing the data transfer clock rate. The other is widening transfer data width. The first approach appears in high-bandwidth memory, such as high-performance DDR memory and HBM memory. Currently, the peak performance of such memory is around 256GB/s\cite{b28}. An example of the second approach is in Interposer, where the number of connections is in the 1000’s \cite{b10}.

The second challenge for AI chips is power consumption. Training an AI model emits more than five times the lifetime emissions of an average car \cite{b12}. Generally, power consumption comes mainly from computing units or data transfer units. AI chips must handle not only computationally intensive but also data transfer-intensive tasks. In addition to algorithm optimization, common approaches to reducing power consumption include using more advanced IC fabrication technology and reducing capacitance load on data transfer paths. To lower the power consumption of data transfer between DRAM \cite{b10} and AI chips, high-bandwidth memory (HBM) \cite{b11} and package substrate routing, like Interposer, are used. Yet even with these advanced technologies, data transfer power consumption is still over 0.5mW/Gbps. \cite{b18}

The third challenge for AI chips is fast memory capacity \cite{b14}.  Fast memory is referred to as on-chip or near-chip memory, which provides data quickly without throttling the processing units. Fast memory is generally in the form of on-chip SRAM. SRAM has the advantage of fast access time, but its size limits memory capacity. As AI is applied to more complex problems, the number of parameters of neural network models grows exponentially. The largest NLP model to date has 170 billion parameters. It is crucial to keep as much  parameters as possible in fast memory to avoid performance degradation. Currently, no AI chip can hold that many parameters. Most AI chips on the market typically have a memory capacity of just 50 MB, which leaves a large gap between current memory capabilities and memory demands. The insufficiency of fast memory capacity will only continue to get worse if nothing is done to address this problem.

The main goal of our approach is to overcome memory bandwidth limitations, to reduce power, and increase memory capacity. We achieve this by optimizing integration, architecture, and memory choice. The component technologies for our Sunrise chip include Heterogeneous Integration Technology on Chip (HITOC), single form memory (UNIMEM), and architecture specifically optimized for HITOC and UNIMEM. 

\section{HITOC}
\begin{figure}[htbp]
\centerline{\includegraphics[width=0.8\linewidth]{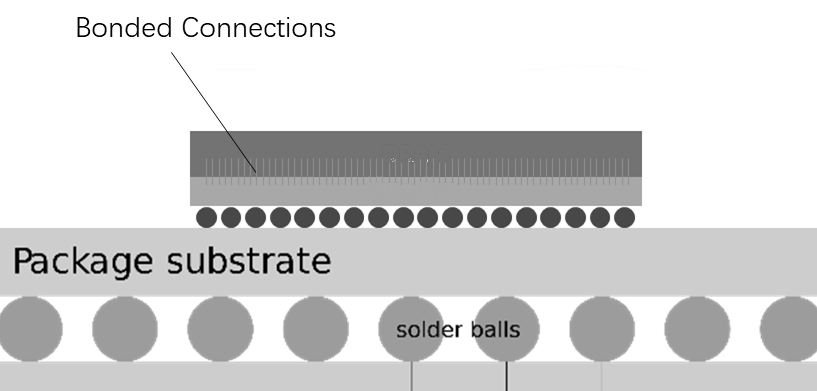}}
\caption{Connection between two wafers with HITOC}
\label{fig:HITOC}
\end{figure}

All AI chips are currently fabricated on a single wafer.  Sunrise chip is partitioned into separately fabricated logic and memory wafers. These two wafers are bonded face-to-face with a hybrid bonding process \cite{b8} \cite{b16} using Cu damascene-patterned surfaces, as shown in Figure \ref{fig:HITOC}. External IOs are brought out to the backside of CMOS wafer with TSV (Through-Silicon Via) for bonding. Circuits on different wafers operate together and communicated through wires running between two wafers. We call this approach Heterogeneous Integration Technology on Chip (HITOC). 

\begin{figure}[htbp]
\centerline{\includegraphics[width=0.8\linewidth]{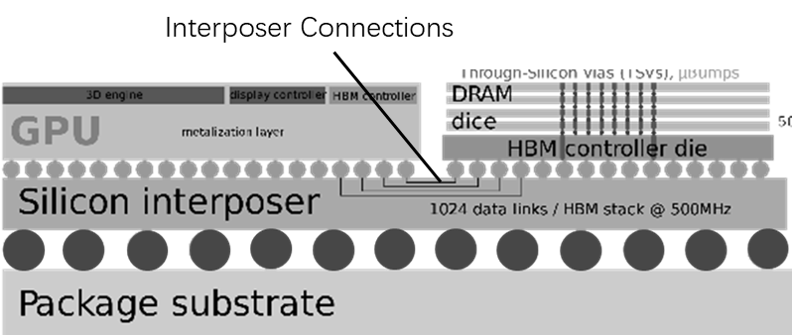}}
\caption{Interposer to connect two chips}
\label{fig:Interposer}
\end{figure}

By fabricating logic and memory wafers separately, both the logic process and memory process are optimized independently. In the logic process, the transistor threshold voltage is low, and the number of metal layers is large. The opposite is true in the memory process. The physical structure of basic elements in the logic process and the memory process differ. Integrating logic and memory processes thus results in a design that favors one over the other, so we choose to separate the two. Furthermore, with logic and memory on separate wafers, there are more compute elements and memory elements in the same area when compared to a single wafer. By separating logic and memory, we achieve better electrical characteristics and overall chip characteristics as well as higher computational performance and memory capacity.
HITOC is one form of 3D IC. There are other forms of 3D IC that are widely used. One is Interposer.  Interposer \cite{b29} is to connect two chips through metal lines on the substrate, as in Figure \ref{fig:Interposer}. Interposer connections are much denser  compared to connections between different package of chips or bonding wire connections between chips in the same package. 

\begin{figure}[htbp]
\centerline{\includegraphics[width=0.8\linewidth]{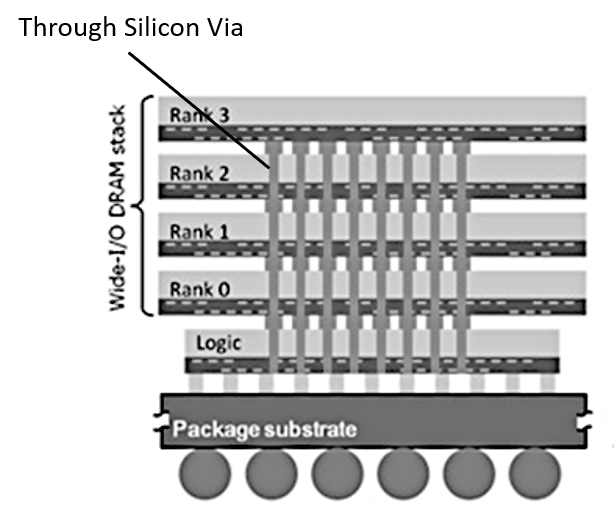}}
\caption{Through Silicon Via connect multiple chips on a stack}
\label{fig:TSV}
\end{figure}

Another approach is Through-Silicon Via. TSV is a direct vertical connection between different levels of a chip. It consists of a conducting via which passes through the silicon substrate and connects the two sides of the wafer [Fig. \ref{fig:TSV}]. Typically, the interplane via is etched and filled with metal, such as tungsten (W) or copper (Cu). Connections between chips with are  denser compared to Interposer. TSV is commonly used in HBM (high bandwidth memory).

Compared to Interposer and TSV, HITOC is even denser because spacing between connections are smaller.  With Interposer, connections are in one dimension between two chips placed on the same surface. With TSV and HITOC, connections are in two dimensions between two chips stacked together.  As shown in Table \ref{table:wire},  wire pitch difference among all three approach directly affects wire density. It ultimately makes the difference in bandwidth which is in proportion to number of connections \cite{b1} \cite{b9} \cite{b10}.

\begin{table}[htbp]
\caption{Data path comparisons of Interposer, TSV, and HITOC}
\begin{center}
\begin{tabular}{|c|c|c|c|}
\hline
&Interposer&TSV&HITOC\\
\hline
Wire Pitch (um)&11.5&9.2×9.2&1×1\\
\hline
Wire Density ($/mm^2$)&86&$1.2 \times 10^4$& $1 \times 10^6$\\
\hline
Bandwidth (TB/s)&0.086&1.2&100\\
\hline
\multicolumn{4}{l}{}$^a$100mm2 die with 1\% of connection area for TSV and wafer stacking, \\
\multicolumn{1}{l}{}1GHz I/O frequency.
\end{tabular}
\label{table:wire}
\end{center}
\end{table}

In that table, we conservatively assume the clock frequencies for data transfer are the same for comparison purposes. But clock frequency runs faster at HITOC and TSV. Data paths are  shorter in HITOC than in Interposer and TSV. Shorter data paths have smaller capacitance loading. As a result, power consumption is only 0.02pJ/b for HITOC while 2.17 pJ/b and 0.55 pJ/b for Interposer and TSV respectively.  With smaller capacitance loading, data transfer runs at higher frequency, and power consumption is lower. 

With shorter and denser connection, HITOC delivers, among the three approaches, higher data transfer rate and lower power consumption.

\section{UNIMEM: A SINGLE MEMORY SOLUTION}
To simplify the system and circumvent the need for conventional CPU-cache-memory architecture \cite{b2}, Sunrise chip uses only the DRAM without SRAM cache. We call this approach UniMem, as we use only a single form of memory for the whole chip.  We choose to use DRAM over SRAM, since DRAM has a higher density with a cell size of 6-12 $F^2$ compared to SRAM’s cell size of 140 $F^2$ \cite{b25}. However, DRAM has a read/write latency that is around 50-90 times slower than SRAM’s \cite{b25}.

\begin{figure}[htbp]
\centerline{\includegraphics[width=\linewidth]{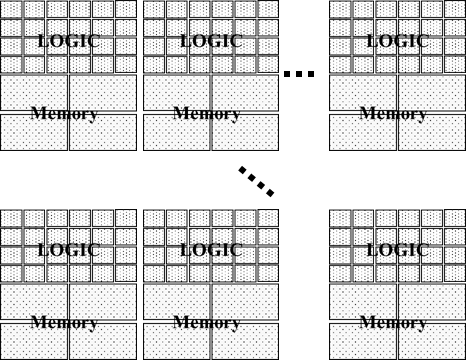}}
\caption{Localized dedicated memory for logic units}
\label{fig:local}
\end{figure}

To counteract  DRAM’s slow latency, multiple localized DRAM units are pooled together to supply data to logic units [Fig. \ref{fig:local}].  Memory access load is shared amongst DRAM arrays in the pool. 

The computation sequence is rearranged such that parameters are reused. We adopted weight stationary data flow \cite{b3}. Operations on the same weights are grouped so that access to weight data from memory is minimized. 

Data is broadcast and shared: in our weight-stationary systolic architecture, feature data and results move. Input feature data is broadcast to all Vector Processing Units (VPU). Each VPU computes and generates output channels independently from other cores. The results are sent back to a central memory pool. Then, the VPU performs all the operations necessary to generate a result. All intermediate data are localized in VPU's, and no exchange of such data occurs with any other units.

With localized DRAM array pooling, and maximizing data movement, Sunrise chip overcomes slow DRAM latency and deliver high computation performance.
\section{CHIP ARCHITECTURE}
\begin{figure}[htbp]
\centerline{\includegraphics[width=\linewidth]{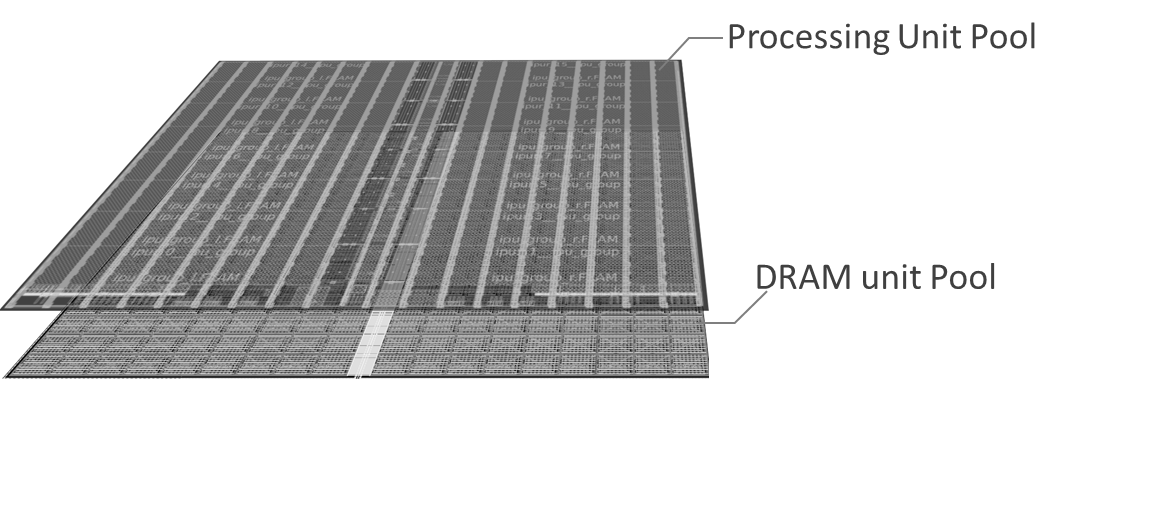}}
\caption{HTOC with Logic wafer and DRAM wafer}
\label{fig:stack}
\end{figure}

With HITOC, we have two wafers, logic wafer, and memory wafer, bonded together [Fig. \ref{fig:stack}]. On the logic wafer, we have pools of processing units.  Underneath the logic pool on the other wafer are pools of DRAM arrays.

\begin{figure}[htbp]
\centerline{\includegraphics[width=\linewidth]{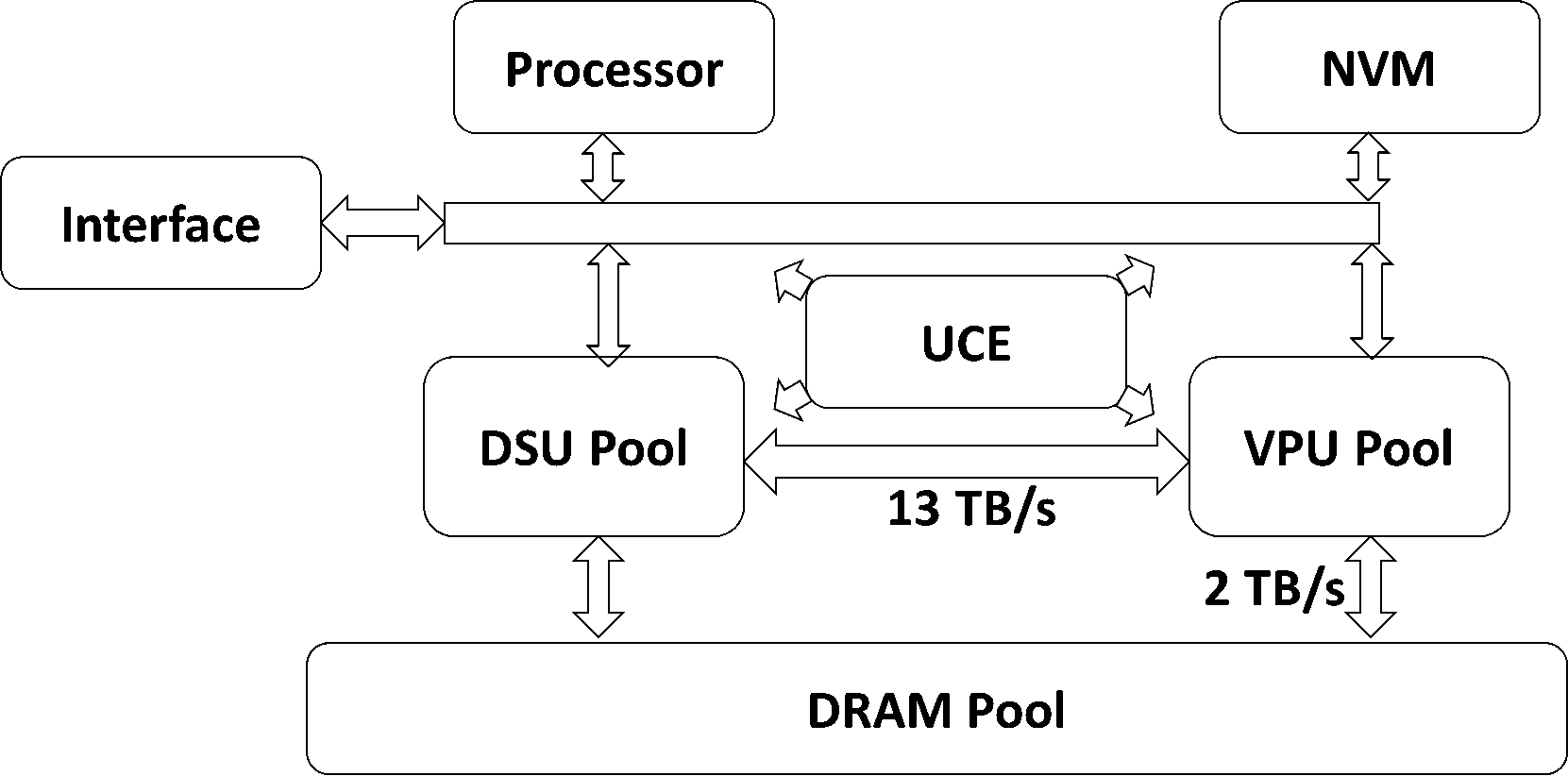}}
\caption{Top level architecture}
\label{fig:overview}
\end{figure}

Logic wafer consists of mainly logic units, and control units that include processor and unified control engine (UCE) [Fig. \ref{fig:overview}].

There are two types of logic units: data serving unit (DSU) and vector processing unit (VPU). VPU's perform computation on data. DSU's serve data to VPU. Each DSU and VPU has their own multiple DRAM arrays directly bonded below the units from the DRAM wafer. All VPU's and DSU's form their respective pool. The overall bandwidth between DSU/VPU pool and DRAM pool is 1.8TB/s. 

Feature data are stored in the DRAM of the DSU pool and are sent to the VPU pool for computation. The results are sent back to the DSU pool. The data bandwidth between the DSU pool and the VPU pool is 13 TB/s. This high bandwidth ensures that data transfer between DSU and VPU is not a bottleneck.

Because memory bandwidth is abundant in Sunrise chip, we choose to use vectors instead of tensors as the basic computational data unit. This allows us to optimize for better computational performance on sparse tensors.  

With extremely high data bandwidth on Sunrise chip, synchronization of all modules is challenging. On this chip, all data flow and module operations are centrally controlled by a single unit called the Unified Control Engine (UCE). It consists of modules such as a Direct Memory Access controller (DMA), data path multiplexer controllers, and function selector. All modules are fully configurable to implement different neural networks. 

There is a proprietary 13-bit processor on Sunrise chip.  It mainly controls high-level tasks such as data batch movement and UCE configuration.

To minimize yield loss due to defects in memory, our DRAM PHY is capable of DRAM repair. Before shipment, DRAM is tested, and defects are recorded in non-volatile memory (NVM). During chip power-up, the defect information is retrieved, and repairs are applied to DRAM arrays. 

There are two chip interfaces. One is a standard SPI interface, and the other is a proprietary high-speed-port (HSP) interface. SPI is for the host to transfer commands to the chip. The HSP interface is for data transfer with a transfer rate of 200MB/s.    

\begin{figure}[htbp]
\centerline{\includegraphics[width=\linewidth]{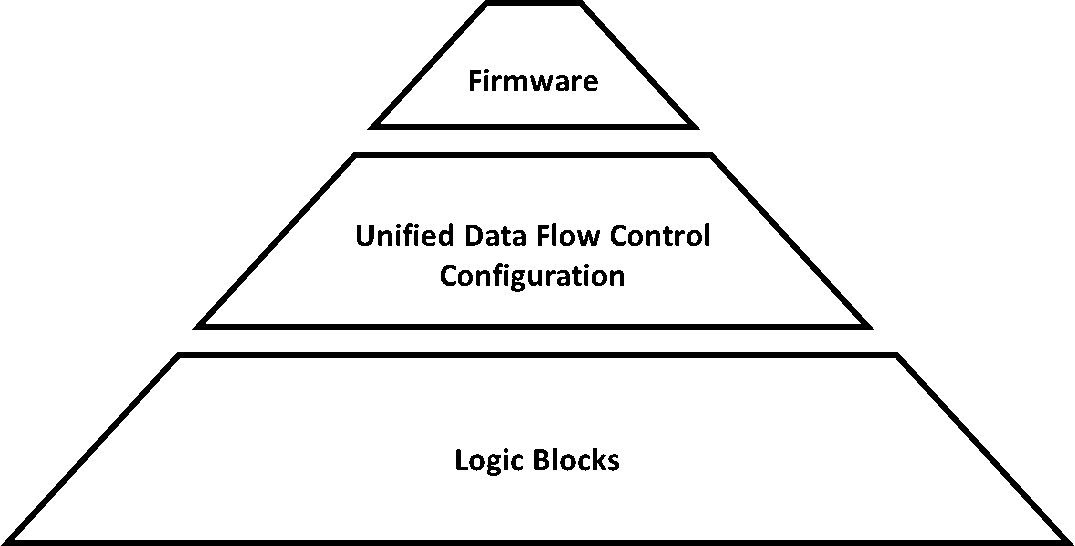}}
\caption{Implementation layers}
\label{fig:tiers}
\end{figure}

Sunrise chip has three implementation layers: logic blocks, unified data flow control configuration, and firmware [Fig. \ref{fig:tiers}]. Logic blocks consist of primitive functional blocks and configurable modules. The unified data flow control configuration dictates how the configurable module functions.  It also initiates predetermined sequences of operations. The top tier is the firmware. Firmware mainly modifies operation register values, changes configurations, or calls out configurations. The firmware also initiates large operations whose sequence is controlled by configuration.  It is also responsible for host and chip communication.  With this three-tier architecture, Sunrise chip implements a wide range of neural networks through a combination of firmware and configuration.
\section{Results}
\begin{figure}[htbp]
\centerline{\includegraphics[width=0.5\linewidth]{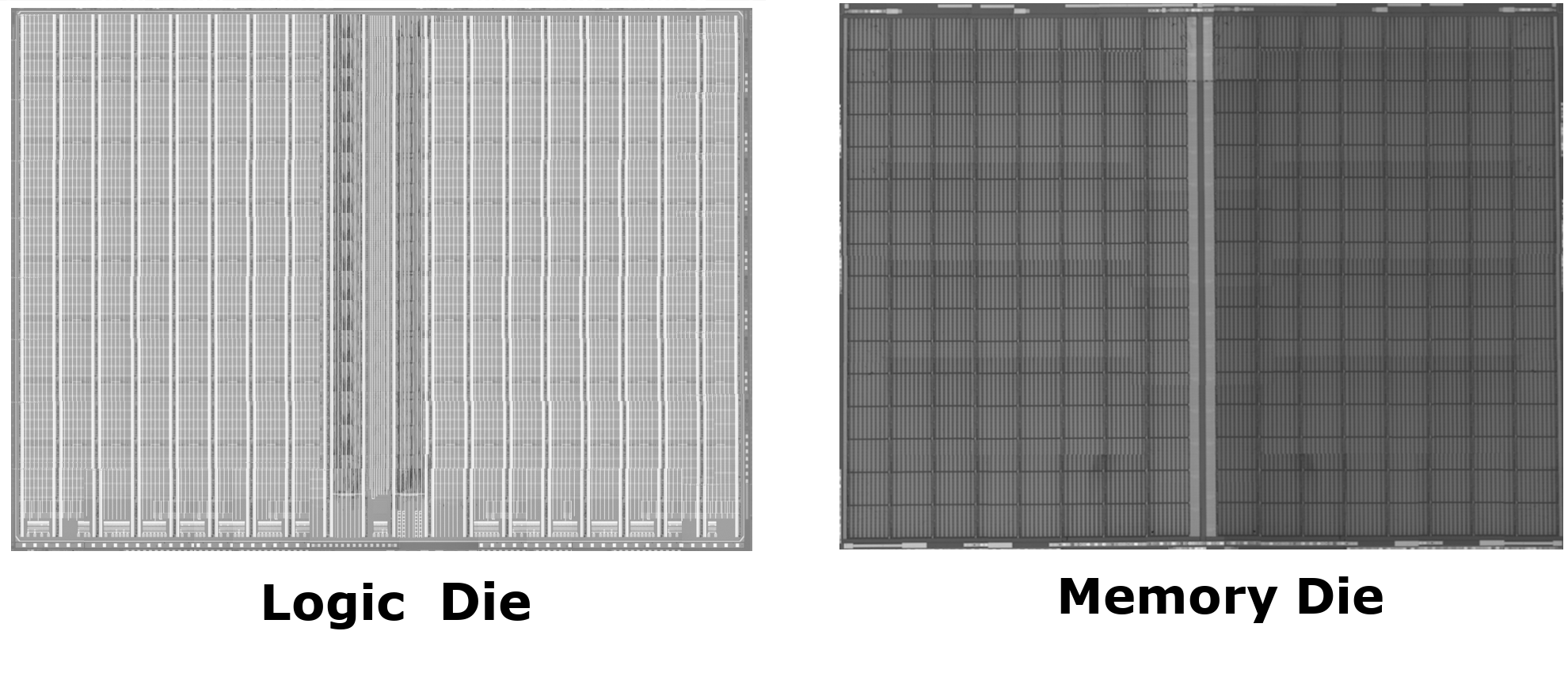}}
\caption{Dies that make up a chip}
\label{fig:die}
\end{figure}

We fabricated Sunrise chip with a 40nm CMOS process and 38nm DRAM process. The chip consists of one die from logic wafer and one die from memory wafer [Fig. \ref{fig:die}]. There are 32,768 MAC on the chip, with a die size of 110 $mm^2$ (12.4 mm $\times$ 8.8 mm) The chip has a high speed data interface with a transfer rate of 200MB/s and an SPI command interface. It has a peak performance of 25 TOPS.  It performances inference of 1500 images per second with ResNet50 model.  Internal memory bandwidth is 1.8TB/s. Internal memory capacity is 4.5Gb.  The typical power consumption is 12W. Its temperature operating range is between -40 \degree C and 85 \degree C. 

We compare Sunrise chip with three other leading AI chips, whose information is publicly available. Table \ref{table:comp} contains key metrics and specifications for Sunrise as well as the other chips, referred to as Chip A \cite{b19}, B \cite{b20}, and C \cite{b21}.
\begin{table}[htbp]
\caption{Benchmark Results for Sunrise}
\begin{center}
\begin{tabular}{|c|c|c|c|c|}
\hline
&\textbf{Sunrise}&\textbf{Chip A}&\textbf{Chip B}&\textbf{Chip C}\\
\hline
Process & 40nm & 16 nm & 12 nm & 7 nm\\
\hline
Die Size ($mm^2$) & 110 & 800 & 709 & 456\\
\hline
Peak Performance (TOPS) & 25 & 122 & 125 & 512\\
\hline
Memory Capacity (MB) & 560 & 300 & 190 & 32\\
\hline
Power Consumption (W) & 12 & 120 & 280 & 350\\
\hline
Memory Bandwidth (TB/s) & 1.8 & 45 & no data & 3\\
\hline

\end{tabular}
\label{table:comp}
\end{center}
\end{table}

Each chip has a different die size. We remove this factor by normalizing by die size and compare the chips on the following benchmarks [Tab. \ref{table:die2die}]:
\begin{itemize}
    \item Peak performance in TOPS/mm$^2$ (trillion operations per second per square millimeter).  Computation performance is higher with more computation units. Number of computation units increases with larger die.  We normalize performance by unit die area for the true performance comparison.
    \item Memory Bandwidth on unit die area in MB/s/mm$^2$. Memory bandwidth is higher with more localized memory arrays. We normalize the memory bandwidth by unit area.
    \item Energy efficiency as a ratio of performance to power consumption, in TOPS/W. 
    \item Memory Capacity per unit area in MB/mm$^2$
    \item Cost. Cost comparison includes standard non-recurrence expenses, die cost per units area, and for application purpose, cost per unit performance. 
\end{itemize}

\begin{table}[htbp]
\caption{Die-to-Die Benchmark Comparisons}
\begin{center}
\begin{tabular}{|c|c|c|c|c|}
\hline
 & \textbf{Peak}
 & \textbf{Memory}
 & \textbf{Memory}
 & \textbf{Energy}\\
 & \textbf{Performance}$^{\mathrm{a}}$
 & \textbf{Bandwidth}$^{\mathrm{b}}$
 & \textbf{Capacity}$^{\mathrm{c}}$
 & \textbf{Efficiency}$^{\mathrm{d}}$\\
\hline
SUNRISE (40nm) & 0.23 & 16.3 & 5.11 & 2.08 \\
Chip A (16nm) & 0.15 & 56.2 & 0.38 & 1.02\\
Chip B (12nm) & 0.18  & no data& 0.27 & 0.45\\
Chip C (7nm) & 1.12 & 6.6 & 0.07 & 1.46\\
\hline
\multicolumn{2}{l}{$^{\mathrm{a}}$Measured in TOPS/$\text{mm}^2$.}
& \multicolumn{2}{l}{$^{\mathrm{b}}$Measured in MB/s/$\text{mm}^2$.} \\
\multicolumn{2}{l}{$^{\mathrm{c}}$Measured in MB/$\text{mm}^2$.}
& \multicolumn{2}{l}{$^{\mathrm{d}}$Measured in TOPS/W.}
\end{tabular}
\label{table:die2die}
\end{center}
\end{table}
Sunrise chip outperforms on two of the four metrics, memory capacity and energy efficiency [Tab. \ref{table:die2die}]. Its peak performance is below chip C's. This is understandable considering that Sunrise is fabricated on 40nm process, a process four generations behind that of chip C.

Sunrise memory bandwidth is below chip A's. Chip A has large amount of fast SRAM on chip. SRAM enables high memory bandwidth. However, SRAM memory bit units are large and take up a large portion of the die area. This reduces memory capacity and leaves smaller die area for computation units. As a result, chip A lags behind Sunrise in both memory capacity and performance.

Sunrise takes a different approach to balance trade-offs between memory and perfomrance. Replacing SRAM with DRAM leads to large memory capacity and leaves more die area for computation units. With HITOC and UNIMEM technology, this architecture not only overcomes slow DRAM latency but also has sufficient memory bandwidth to support high performance. HITOC cuts down data transfer energy consumption. UNIMEM removes SRAM cache, and thus the energy consumption associated with it. Both factors make Sunrise the most energy efficient amongst the other chips in the table.

\begin{table}[htbp]
\caption{Cost Comparison in USD}
\begin{center}
\begin{tabular}{|c|c|c|c|}
\hline
 & \textbf{NRE}
 & \textbf{Die Cost}
 & \textbf{Cost per TOPS}\\
\hline
SUNRISE (40nm) & $2.2\times 10^6$ &  11 & 0.43 \\
Chip A (16nm) & $7.2\times 10^6$ &  617 & 2.47 \\
Chip B (12nm) & $15\times 10^6$  &  296 & 1.19 \\
Chip C (7nm) & $24\times 10^6$  &  336 & 0.66 \\
\hline
\end{tabular}
\label{table:cost}
\end{center}
\end{table}

In Table \ref{table:cost}, we compare non-recurrence expense (NRE) and die costs. NRE mainly consists of process mask cost. Although the exact die cost of the other chips are not published, we estimate their die cost based on die size, wafer cost from major foundries, and expected yields. We also include the cost for delivering the same performance. It is expected that 40nm process delivers the lowest cost. Normally, a more advanced process delivers better cost-to-performance ratio. However, our chip delivers the best cost-to-performance ratio even with 40nm process. This is directly due to our chip architecture. 

Combining all the discussed metrics, our Sunrise chip overall outperforms other leading AI chips despite its less advanced fabrication process. The HITOC and UNIMEM technology incorporated in this chip allows us move to a new and more optimal architecture. The key benefit is being able to achieve and surpass the performance of more advanced fabrication processes with less expensive fabrication.

\section{Projection}

The AI chips in comparisons are fabricated with different processes. To compare the architecture effectively, we normalize each chip to a 7nm CMOS process and a 1y DRAM process, based on factors such as density, transistor performance, and power reduction. The factors are derived from the parameters of the CMOS process (Table \ref{table:tsmc}) \cite{b22} \cite{b23} and the DRAM process (Table \ref{table:samsung})\cite{b24} of leading foundries.

\begin{table}[htbp]
\caption{CMOS process parameters}
\begin{center}
\begin{tabular}{|c|c|c|c|}
\hline
&\textbf{Density}&\textbf{Performance}&\textbf{Power}\\
&\textbf{Ratio}&\textbf{Improvement}&\textbf{Reduction}\\
\hline
28 nm vs. 40 nm& 2 & 45\% & 40\%\\
\hline
16 nm vs. 28 nm& 2 & 35\% & 55\%\\
\hline
12 nm vs. 16 nm& 1.2 & 28\% & 35\%\\
\hline
10 nm vs. 16 nm& 2 & 15\% & 35\%\\
\hline
7 nm vs. 10 nm& 1.65 & 22\% & 54\%\\
\hline
\end{tabular}
\label{table:tsmc}
\end{center}
\end{table}

\begin{table}[htbp]
\caption{DRAM density}
\begin{center}
\begin{tabular}{|c|c|c|}
\hline
3x nm Process & 1x nm Process & 1y nm Process\\
\hline
0.04 Gb/mm$^2$& 0.189 Gb/mm$^2$ & 0.237 Gb/mm$^2$ \\
\hline
\end{tabular}
\label{table:samsung}
\end{center}
\end{table}

As seen in Table \ref{table:tsmc}, density is improved with each generation of process. More transistor computing units can be packed into the same die area. This not only improves device performance but also increases power consumption by the same factor. With more advanced process generation, one can choose high performance process to get performance gains, or choose lower power process to get better power efficiency. When we project the parameters of each chip in our normalization calculations, we use performance improvement parameters under the condition that power consumption is within the common range as seen in ASIC chips. Otherwise, we use power reduction parameters. 

With all the chips normalized to 7nm, Sunrise chip architectures surpass all other three chips in all benchmarks (Table \ref{table:projection}).
\begin{table}[htbp]
\caption{Benchmark Comparisons Normalized to 7nm process}
\begin{center}
\begin{tabular}{|c|c|c|c|c|}
\hline
 & \textbf{Peak}
 & \textbf{Memory}
 & \textbf{Memory}
 & \textbf{Energy}\\
 & \textbf{Performance}$^{\mathrm{a}}$
 & \textbf{Bandwidth}$^{\mathrm{b}}$
 & \textbf{Capacity}$^{\mathrm{c}}$
 & \textbf{Efficiency}$^{\mathrm{d}}$\\
\hline
SUNRISE & 7.58 & 216 & 30.3 & 50.10 \\
Chip A & 0.86 & 122 & 1.50 & 5.38\\
Chip B & 0.19 & no data &  0.90 & 0.83\\
Chip C & 1.12 & 6.6 & 0.07 & 1.46\\
\hline
\multicolumn{2}{l}{$^{\mathrm{a}}$Measured in TOPS/$\text{mm}^2$.}
& \multicolumn{2}{l}{$^{\mathrm{b}}$Measured in MB/s/$\text{mm}^2$.} \\
\multicolumn{2}{l}{$^{\mathrm{c}}$Measured in MB/$\text{mm}^2$.}
& \multicolumn{2}{l}{$^{\mathrm{d}}$Measured in TOPS/W.}
\end{tabular}
\label{table:projection}
\end{center}
\end{table}

Although Sunrise chip is on a 40nm process, its performance per unit area exceeds that of two competing chips at 12nm and 16nm. It exceeds all three chips after normalized by process node. With the architecture of Sunrise, one can choose to either use a less expensive process to achieve the same performance as other chips or use current processes to get better performance then other chips. 

As shown in Table \ref{table:die2die}, the Sunrise chip has the highest memory bandwidth. We designed just enough bandwidth as needed for chip performance. Table \ref{table:wire} shows that HITOC technology enables extremely high memory bandwidth. Even so, bandwidth after normalization exceeds that of all comparable devices. 

Sunrise chip’s memory capacity is 20 times that of other chips at the 40nm node. When we normalized it to 7nm CMOS and 1Ynm DRAM process, it would reach 20 times the memory capacities of other chips. The gain in memory capacity is mostly a result of replacing all SRAM with DRAM, which has a  density of more than 14 times higher than SRAM \cite{b13}. On a 800 $mm^2$ die, our architecture could reach a storage capacity as high as 24GB. The largest memory capacity on AI chip ever made is 18GB \cite{b17}, which requires a whole wafer. With our architecture, the current memory capacity of an AI wafer can fit onto a single chip.

Sunrise chip is more energy-efficient than all other three chips, even though it is on a less advanced process. Our high energy efficiency is due to the removal of SRAM cache and the close proximity between memory and compute units.

Overall, Sunrise chip meets or exceeds benchmarks of AI chips in the market.  The Sunrise architecture is projected to well exceed all benchmarks if fabricated with an advanced process.

\section{Conclusion}
We fabricated an AI chip, Sunrise, with our HITOC and UNIMEM technology. We developed a special architecture to overcome slow DRAM latency and completely replace SRAM with high-capacity DRAM. Sunrise chip, despite being fabricated on a 40nm process node, matches or exceeds other AI chips with more advanced process technology on the metrics of performance, memory bandwidth, memory capacity, energy efficiency, and cost. Based on the data of current silicon, we project that the same architecture is 7 to 20 times better on all major benchmarks. This architecture breaks the memory wall for AI applications. While our designs are motivated by the demands required by deep neural network training and inference, we believe that these technologies are applicable to chips  used in other applications, such as high performance computing and big data processing.

\end{document}